\documentclass
[preprint,prd,twocolumn,10pt,letterpaper,amsfonts,showpacs,floats]{revtex4}%
\usepackage{amsfonts}
\usepackage{amsmath}
\usepackage{amssymb}
\usepackage{graphicx}%
\begin{document}
\title{Wheeler-De Witt equation for brane gravity}
\author{Pawel Gusin}
\affiliation{University of Silesia, Institute of Physics, ul. Uniwersytecka 4, PL-40007
Katowice, Poland}


\begin{abstract}
We consider the gravity in the system consisting of the
Bogomol'nyi-Prasad-Sommerfield (BPS) D3-brane embedded in the flat background
geometry, produced by the solutions of the supergravity. The effective action
for this system is represented by the sum of the Hilbert-Einstein and DBI
actions. We derive the Wheeler-De Witt equation for this system and obtain
analytical solutions in some special cases. We also calculate tunneling
probability from Planckian size of D3-brane to the classical regime.

\end{abstract}
\maketitle

\section{Introduction}

The discovery of D branes in the string theory have opened the way to the
realization of the universe as a braneworld embedded\ in an ambient space. In
the type IIA, IIB and I strings there are many D branes sustaining gauge and
matter fields, at least one of them should include the Standard Model of
particle physics. This D-brane will be correspond to the our universe. All the
matter and gauge fields are due to the open string sector. The strings from
this sector have the end points constrained to move on the brane. The graviton
and the dilaton are contained in the closed string sector and probe the extra
dimensions. In the regions where the back-reaction of D-brane and stringy
geometry can be ignored, the effective action for the worldvolume in the flat
background is given by the DBI action coupled to the worldvolume gravity. Such
models were used in cosmology in order to explain inflation from the string
theory [1].

The DBI action describes the effects of virtual open strings at the tree
level; it includes the effects of the background geometry and field strengths.
But the non-linear form of the DBI action is inconvenient for quantization.
However, introducing an intrinsic worldvolume metric one can write down the
equivalent action [2]. From the other side one can express the DBI action as
sum of constraints [3-5], where Hamiltonian for the D-brane is the constrain.
In the first case, where DBI action is represented by the intrinsic metric,
one can consider this metric as a source of gravitation and the induced metric
on the worldvolume as a field coupled to the gravity. This case will be
considered. The physical sense of our study is that in the classical
background of 10 dimensional spacetime we consider 4 dimensional quantum
subsystem. In the low-energetic approximation the classical backgrounds are
given by the solutions of the supergravity. One of the best known and well
investigated are $AdS_{k}\times X_{10-k}$ backgrounds (AdS/CFT
correspondence). The other are backgrounds with warped metrices
maintaining\ the four-dimensional Poincare symmetry. The most interesting
backgrounds are responsible for inflation and leading to the de Sitter vacuum
in four dimensions [6].

In this paper we extend the applicability of the models with the DBI action to
the quantum regime. In this regime we have to take into account quantum
effects as well as for the gravity and for the other fields from the string
sector (or low-energetic approximation). These effects can be described by the
Wheeler-De Witt equation. Motivated by the above remarks we consider the
Friedmann-Robertson-Walker (FRW) model coupled to the DBI action. In section 2
we derive a classical action for the considered system. This action is the
base for section 3 where we obtain the Wheeler-De Witt equation and give
solutions of this equation in particular cases. Section 4 is devoted to conclusions.

\section{DBI action and gravity}

The well-known form of DBI action for a D3-brane is (modulo WZ terms):%
\begin{equation}
S_{3}=-T_{3}\int d^{4}xe^{-\phi}\sqrt{-\det(\gamma_{\alpha\beta}+2\pi
\alpha^{\prime}F_{\alpha\beta}+B_{\alpha\beta})}, \tag{2.1}%
\end{equation}
where $\gamma_{\alpha\beta}$ and $B_{\alpha\beta}$ are pull-backs of\ a
background metric $g_{AB}$ and NS field $B_{AB}$ by an embedding field $X$,
respectively. A strength $F_{\alpha\beta}$ of $U\left(  1\right)  $ gauge
field is the intrinsic field on the worldvolume of D3-brane. As it is shown in
[2] the above action can be expressed in the equivalent forms by an intrinsic
auxiliary worldvolume tensor field $h_{\alpha\beta}$. We consider the simplest
case when the fields $F_{\alpha\beta}$ and $B_{\alpha\beta}$ on the
worldvolume vanish. Then the action (2.1) becomes the 4-dimensional Nabu-Goto,
which takes the form:%
\begin{equation}
S_{3}^{\prime}=-\frac{\Lambda T_{3}}{2}\int d^{4}xe^{-\phi}\sqrt{-\det\left(
h_{\alpha\beta}\right)  }\left[  h^{\alpha\beta}\gamma_{\alpha\beta}%
-2\Lambda\right]  , \tag{2.2}%
\end{equation}
where $h_{\alpha\beta}$ is the intrinsic auxiliary metric on the worldvolume
and $\Lambda$ is a constant. We promote the metric $h_{\alpha\beta}$ to the
dynamic degree of freedoms. Thus the metric $h_{\alpha\beta}$ is a field which
is responsible for gravitation on the worldvolume and the metric $\gamma$ on
the worldvolume is considered as a field induced by the background. Hence the
system consists of gravitation on the worldvolume coupled to the fields
originated from the background. Problems of the backreaction in this
consideration are negleced. So the action for this system is obtained from
(2.2) by adding the Einstein-Hilbert term:%
\begin{gather}
S=\frac{m_{P}^{2}}{2}\int d^{4}x\sqrt{-\det\left(  h_{\alpha\beta}\right)
}R\left(  h\right)  -\nonumber\\
\frac{\Lambda T_{3}}{2}\int d^{4}xe^{-\phi}\sqrt{-\det\left(  h_{\alpha\beta
}\right)  }\left[  h^{\alpha\beta}\gamma_{\alpha\beta}-2\Lambda\right]  ,
\tag{2.3}%
\end{gather}
where $m_{P}^{2}=\left(  8\pi G\right)  ^{-1}$. Next we apply the ADM
construction for this system. In this construction the worldvolume $M$ is the
product: $\mathbf{R}^{1}\times\Sigma_{3}$ where $\Sigma_{3}$ is the
3-dimensional space-like slice of $M$. Then the metric $h_{\alpha\beta}$ is
determined by a shift vector $N^{m}$ and a lapse function $N$ as follows:%
\[
h_{00}=-N^{2}+\overline{h}_{mn}N^{m}N^{n},\text{ \ }h_{0m}=\overline{h}%
_{mn}N^{n},\text{ \ }h_{mn}=\overline{h}_{mn},
\]
where $\overline{h}_{mn}$ is the intrinsic metric on $\Sigma_{3}$. The matrix
$\left(  h^{\alpha\beta}\right)  $ has the entries:%
\[
h^{00}=-1/N^{2},\text{ }h^{0m}=N^{m}/N^{2}\text{, }h^{mn}=\overline{h}%
^{mn}-N^{m}N^{n}/N^{2}.
\]
Using the above relations we get:%
\[
h^{\alpha\beta}\gamma_{\alpha\beta}=-\frac{1}{N^{2}}\left(  \gamma_{00}%
+N^{m}N^{n}\gamma_{mn}-2N^{m}\gamma_{0m}\right)  +\overline{h}^{mn}\gamma
_{mn}.
\]
The action (2.3) in the comoving coordinates ($N^{m}=0$) $h_{mn}=\overline
{h}_{mn}$ and for the FRW metric :%
\begin{equation}
ds^{2}=-N^{2}dt^{2}+a^{2}\left(  t\right)  \left[  \frac{dr^{2}}%
{1-\widetilde{k}R^{2}}+R^{2}\left(  d\theta^{2}+\sin^{2}\theta d\phi
^{2}\right)  \right]  \tag{2.4}%
\end{equation}
results in:%
\begin{gather}
S=\frac{m_{P}^{2}}{2}\int_{\Sigma_{3}}d\mu\int_{\mathbf{R}^{1}}dta^{3}N\left[
\frac{\widetilde{k}}{a^{2}}-\left(  \frac{\overset{\cdot}{a}}{Na}\right)
^{2}\right]  -\nonumber\\
\frac{\Lambda T_{3}}{2}\int_{\mathbf{R}^{1}}dt\int_{\Sigma_{3}}d\mu
a^{3}e^{-\phi}N\left[  -\frac{\gamma_{00}}{N^{2}}+\frac{1}{a^{2}}h^{mn}%
\gamma_{mn}-2\Lambda\right]  , \tag{2.5}%
\end{gather}
where
\[
d\mu=\frac{6R^{2}\sin\theta}{\sqrt{1-\widetilde{k}R^{2}}}drd\phi d\theta
\]
and%
\[
h^{mn}\gamma_{mn}=\left(  1-\widetilde{k}R^{2}\right)  \gamma_{RR}+\frac
{1}{R^{2}}\gamma_{\theta\theta}+\frac{1}{R^{2}\sin^{2}\theta}\gamma_{\phi\phi
}.
\]
For our purpose we assume that $\Sigma_{3}$ has $O\left(  4\right)  $ symmetry
and the metric $h_{mn}$ is the metric on $S^{3}$ so $\widetilde{k}=+1$. Hence
the action reads:%
\begin{gather}
S=6\pi^{2}m_{P}^{2}\int_{\mathbf{R}^{1}}dta^{3}N\left[  \frac{1}{a^{2}%
}-\left(  \frac{\overset{\cdot}{a}}{Na}\right)  ^{2}\right]  +\nonumber\\
6\pi^{2}\Lambda T_{3}\int_{\mathbf{R}^{1}}dta^{3}e^{-\phi}N\left[
\frac{\gamma_{00}}{N^{2}}+2\Lambda\right]  -\nonumber\\
\frac{\Lambda T_{3}}{2}\int_{\mathbf{R}^{1}}dta\int_{S^{3}}d\mu e^{-\phi
}NTr(\gamma), \tag{2.6}%
\end{gather}
where%
\[
Tr\left(  \gamma\right)  =h^{mn}\gamma_{mn}.
\]
The function $N(t)$ can be arbitrarily chosen by a redefinition of time thus
we will use the gauge $N=1$.

This system we put in the flat background produced by $N$ coincident BPS
$Dp$-branes. The metric $g_{MN}$, the dilaton $\phi$ and RR-field $C$ for this
background have the form [7]:%
\begin{gather}
ds_{10}^{2}=g_{MN}dX^{M}dX^{N}=H_{p}^{-1/2}\eta_{\mu\nu}dX^{\mu}dX^{\nu}%
+H_{p}^{1/2}dX_{I}dX^{I},\text{ }\tag{2.7}\\
\text{\ }\left(  \mu,\nu=0,1...,p\text{ and }I=p+1,...,9\right)  ,\nonumber
\end{gather}%
\begin{equation}
e^{2\phi}=H_{p}^{\left(  3-p\right)  /2}, \tag{2.8}%
\end{equation}%
\begin{equation}
C=\left(  H_{p}^{-1}-1\right)  dX^{0}\wedge...\wedge dX^{p}, \tag{2.9}%
\end{equation}
where $H_{p}$ is the harmonic function of the transverse coordinates ($X_{I}$)
to the world-volume:%
\begin{equation}
H_{p}=1+\frac{Ng_{s}}{r^{7-p}},\text{ \ \ (}r=\left(  X_{I}X^{I}\right)
^{1/2}\text{ )} \tag{2.10}%
\end{equation}
and $\left(  \eta_{\mu\nu}\right)  =diag\left(  -1,+1,...,+1\right)  $.

Here we consider backgrounds produced by $Dp$-branes with $p>3$. We also use
gauge freedom choice for an embedding field $X$ in the form (so called static
gauge):%
\begin{equation}
X\left(  x\right)  =\left(  t,x^{1},...,x^{3},X^{4}\left(  t\right)
,...,X^{9}\left(  t\right)  \right)  . \tag{2.11}%
\end{equation}
In this gauge the part of action related to RR field $C$ vanishes. The induced
metric in this embedding is:%
\[
\gamma_{00}=H_{p}^{-1/2}\left(  -1+\overset{\cdot}{X^{i}}\overset{\cdot}%
{X_{i}}\right)  +H_{p}^{1/2}\overset{\cdot}{X^{I}}\overset{\cdot}{X_{I},}%
\]%
\[
\gamma_{mn}=H_{p}^{-1/2}\delta_{mn},
\]
where $i=4,...,p$. The integral on $S^{3}$ in the last term of (2.6) gives:%
\[
\int_{S^{3}}d\mu e^{-\phi}Tr(\gamma)=12\pi^{2}\left(  1+2+\frac{1}{2}%
\ln2\right)  H_{p}^{\left(  p-5\right)  /4}.
\]
Thus the considered action (2.6) takes the form:%
\begin{gather}
S=6\pi^{2}\int_{\mathbf{R}^{1}}dt\left[  m_{P}^{2}a^{3}\left(  \frac{1}{a^{2}%
}-\left(  \frac{\overset{\cdot}{a}}{a}\right)  ^{2}\right)  \right.
-\nonumber\\
\left.  \Lambda T_{3}\gamma aH_{p}^{\left(  p-5\right)  /4}+L\right]  ,
\tag{2.12}%
\end{gather}
where $L$ is given by the formula:%
\begin{equation}
L=\Lambda T_{3}a^{3}\left[  -H_{p}^{\left(  p-5\right)  /4}\left(
1-\overset{\cdot}{X^{i}}\overset{\cdot}{X_{i}}-H_{p}\overset{\cdot}{X^{I}%
}\overset{\cdot}{X_{I}}\right)  +2\Lambda\right]  \tag{2.13}%
\end{equation}
and $\gamma=1+2+\frac{1}{2}\ln2=3.3466$. In the directions spanned by $X_{i}$
and $X_{I}$ we introduce spherical coordinates:%
\[
X_{i}=\varrho f_{i},\text{ \ }X_{I}=rh_{I},
\]
where $f_{i}f^{i}=h_{I}h^{I}=1$. In the generic case the considered system
(the probed $D3$-brane) has a non-trivial angular momentum in the transverse
directions ($X_{I}$) and non-trivial angular momentum in the directions
transverse ($X_{i}$) to $D3$-brane but longitudinal to the background branes.
For the simplicity, we consider the non-rotating case: $\overset{\cdot}{f}%
_{i}=\overset{\cdot}{h}_{I}=0$. Thus, the eq. (2.13) takes the following form:%
\[
L=\Lambda T_{3}a^{3}\left[  -H_{p}^{\left(  p-5\right)  /4}\left(
1-\overset{\cdot}{\varrho^{2}}-H_{p}\overset{\cdot}{r^{2}}\right)
+2\Lambda\right]  .
\]
Then the three dynamic fields $\left(  a,\varrho,r\right)  $ span
minisuperspace, and the action is:%
\begin{gather}
S=6\pi^{2}\int_{\mathbf{R}^{1}}dt\left[  -m_{P}^{2}a\overset{\cdot}{a}%
^{2}+\right. \nonumber\\
\left.  \Lambda T_{3}a^{3}H_{p}^{\left(  p-5\right)  /4}\left(  \overset
{\cdot}{\varrho^{2}}+H_{p}\overset{\cdot}{r^{2}}\right)  -\widetilde{U}\left(
a,r\right)  \right]  , \tag{2.14}%
\end{gather}
where a potential $\widetilde{U}$ is given by:%
\begin{equation}
\widetilde{U}\left(  a,r\right)  =a^{3}\Lambda T_{3}\left(  H_{p}^{\left(
p-5\right)  /4}-2\Lambda\right)  -a\left(  m_{P}^{2}+\Lambda T_{3}\gamma
H_{p}^{\left(  p-5\right)  /4}\right)  . \tag{2.15}%
\end{equation}
The metric $\widetilde{G}_{\Sigma\Phi}$ on the minisuperspace is taken from
(2.14):%
\begin{equation}
\left(  \widetilde{G}_{\Sigma\Phi}\right)  =\left(
\begin{array}
[c]{ccc}%
-m_{P}^{2}a & 0 & 0\\
0 & \Lambda T_{3}a^{3}H_{p}^{\left(  p-5\right)  /4} & 0\\
0 & 0 & \Lambda T_{3}a^{3}H_{p}^{\left(  p-1\right)  /4}%
\end{array}
\right)  . \tag{2.16}%
\end{equation}
The equation of motion for $\varrho$ gives the following relation:%
\[
\frac{d}{dt}\left[  a^{3}H_{p}^{\left(  p-5\right)  /4}\overset{\cdot}%
{\varrho}\right]  =0.
\]
In this way one obtains:%
\[
\overset{\cdot}{\varrho}=Ja^{-3}H_{p}^{\left(  5-p\right)  /4},
\]
where $J$ is a constant. Thus eq. (2.14) is:%
\begin{equation}
S=6\pi^{2}\int_{\mathbf{R}^{1}}dt\left[  -m_{P}^{2}a\overset{\cdot}{a}%
^{2}+\Lambda T_{3}a^{3}H_{p}^{\left(  p-1\right)  /4}\overset{\cdot}{r^{2}%
}-U\left(  a,r\right)  \right]  \tag{2.17}%
\end{equation}
and a potential $U$ is given by:%
\begin{gather}
U\left(  a,r\right)  =a^{3}\Lambda T_{3}\left(  H_{p}^{\left(  p-5\right)
/4}-2\Lambda\right)  -\nonumber\\
a\left(  m_{P}^{2}+\Lambda T_{3}\gamma H_{p}^{\left(  p-5\right)  /4}\right)
+J^{2}a^{-3}H_{p}^{\left(  5-p\right)  /4}. \tag{2.18}%
\end{gather}
Now, the mini-superspace is reduced and spanned by $a$ and $r$. The metric
$G_{\Sigma\Phi}$ on this space is:%
\begin{equation}
\left(  G_{\Sigma\Phi}\right)  =\left(
\begin{array}
[c]{cc}%
-m_{P}^{2}a & \\
& \Lambda T_{3}a^{3}H_{p}^{\left(  p-1\right)  /4}%
\end{array}
\right)  . \tag{2.19}%
\end{equation}

In the case when $r$ is fixed, the equations for $a$ are obtained from the
Hamiltonian constraint and from the equation of motion for (2.17). These
equations are given by:%
\begin{gather}
\left(  \frac{\overset{\cdot}{a}}{a}\right)  ^{2}=\frac{\Lambda T_{3}}%
{m_{P}^{2}}\left(  H_{p}^{\left(  p-5\right)  /4}-2\Lambda\right)
-\nonumber\\
\frac{1}{a^{2}}\left(  1+\frac{\Lambda T_{3}\gamma}{m_{P}^{2}}H_{p}^{\left(
p-5\right)  /4}\right)  +\frac{1}{a^{6}}\frac{J^{2}}{m_{P}^{2}}H_{p}^{\left(
5-p\right)  /4}, \tag{2.20}%
\end{gather}%
\begin{equation}
\frac{\overset{\cdot\cdot}{a}}{a}=-\frac{1}{m_{P}^{2}}\left[  \Lambda
T_{3}\left(  2\Lambda-H_{p}^{\left(  p-5\right)  /4}\right)  +\frac{2}{a^{6}%
}J^{2}H_{p}^{\left(  5-p\right)  /4}\right]  . \tag{2.21}%
\end{equation}
If one compares these equations to the standard Friedmann equations for a
perfect fluid with an energy density $\rho$, a pressure $p$ and a curvature
parameter $k$:%
\begin{equation}
\left(  \frac{\overset{\cdot}{a}}{a}\right)  ^{2}=\frac{1}{3m_{P}^{2}}%
\rho-\frac{k}{a^{2}}, \tag{2.22}%
\end{equation}%
\begin{equation}
\frac{\overset{\cdot\cdot}{a}}{a}=-\frac{1}{6m_{P}^{2}}\left(  \rho+3p\right)
, \tag{2.23}%
\end{equation}
then one can notice that the energy density $\rho$ is given by:%
\begin{equation}
\rho=3\Lambda T_{3}\left(  H_{p}^{\left(  p-5\right)  /4}-2\Lambda\right)
+\frac{3}{a^{6}}J^{2}H_{p}^{\left(  5-p\right)  /4}, \tag{2.24}%
\end{equation}
the pressure $p$ is expressed as follows:%
\begin{equation}
p=-3\Lambda T_{3}\left(  H_{p}^{\left(  p-5\right)  /4}-2\Lambda\right)
+\frac{3}{a^{6}}J^{2}H_{p}^{\left(  5-p\right)  /4} \tag{2.25}%
\end{equation}
and the curvature parameter $k$ is given by:%
\begin{equation}
k=1+\frac{\Lambda T_{3}\gamma}{m_{P}^{2}}H_{p}^{\left(  p-5\right)  /4}.
\tag{2.26}%
\end{equation}
The first term on r.h.s in (2.24) can be interpreted as a cosmological
constant $\lambda:$%
\begin{equation}
\lambda=3\Lambda T_{3}\left(  H_{p}^{\left(  p-5\right)  /4}-2\Lambda\right)
. \tag{2.27}%
\end{equation}
The state equation for this perfect fluid has the form:%
\begin{equation}
w=\frac{p}{\rho}=-\left(  1-\frac{2J^{2}H_{p}^{\left(  5-p\right)  /4}%
}{\Lambda T_{3}\left(  H_{p}^{\left(  p-5\right)  /4}-2\Lambda\right)
a^{6}+J^{2}H_{p}^{\left(  5-p\right)  /4}}\right)  . \tag{2.28}%
\end{equation}
Hence for $a\rightarrow\infty$ one obtains the following state equation:%
\begin{equation}
w=\frac{p}{\rho}=-1. \tag{2.29}%
\end{equation}
Thus, for big $a$ the worldvolume of D3-brane will be dominated by the perfect
fluid with the negative pressure and the observer fixed to this worldvolume
will see an accelerated expansion.

Let us notice that for $p=5$ the action (2.17) is:%
\begin{gather}
S=6\pi^{2}\int_{\mathbf{R}^{1}}dt\left[  -m_{P}^{2}a\overset{\cdot}{a}%
^{2}+\Lambda T_{3}a^{3}H_{5}\overset{\cdot}{r^{2}}+\right. \nonumber\\
\left.  a\left(  m_{P}^{2}+\Lambda T_{3}\sigma\right)  -J^{2}a^{-3}%
-a^{3}\Lambda T_{3}\left(  1-2\Lambda\right)  \right]  , \tag{2.30}%
\end{gather}
where
\[
H_{5}=1+\frac{Ng_{s}}{r^{2}}.
\]
If one replaces the field $r$ by the following field $\varphi$:%
\begin{equation}
\varphi=\sqrt{Ng_{s}+r^{2}}-Ng_{s}\ln\left(  \frac{Ng_{s}+\sqrt{Ng_{s}+r^{2}}%
}{r}\right)  , \tag{2.31}%
\end{equation}
then the action takes the form:%
\begin{gather}
S=6\pi^{2}\int_{\mathbf{R}^{1}}dt\left[  -m_{P}^{2}a\overset{\cdot}{a}%
^{2}+\Lambda T_{3}a^{3}\overset{\cdot}{\varphi^{2}}+\right. \nonumber\\
\left.  a\left(  m_{P}^{2}+\Lambda T_{3}\gamma\right)  -J^{2}a^{-3}%
-a^{3}\Lambda T_{3}\left(  1-2\Lambda\right)  \right]  . \tag{2.32}%
\end{gather}
Thus for the background produced by $5$-branes the action for $3$-brane is
reduced to the free scalar field $\varphi$ coupled to the scale factor $a$
with the potential $v=-a\left(  m_{P}^{2}+\Lambda T_{3}\gamma\right)
+J^{2}a^{-3}+a^{3}\Lambda T_{3}\left(  1-2\Lambda\right)  $. For $J=0$ the
action (2.32) is the action for the homogeneous and isotropic metric $g$
coupled to the free scalar field with the cosmological constant $\Lambda
T_{3}\left(  1-2\Lambda\right)  $. One can see that the cosmological constant
$\lambda$ is the function of intrinsic parameters describing $3$-brane, namely
the strength $T_{3}$ and the constant $\Lambda$. Thus by a suitable choice of
$\Lambda$ one can obtain the real value (observed) of the cosmological
constant. For $p\neq5$ the cosmological constant is given by (2.27).

\section{The Wheeler-De Witt equation}

In the mini-superspace spanned by the fields $Q=(Q_{1},...,Q_{N})$ the general
form of the Wheeler-De Witt (WD) equation has the form:%
\begin{equation}
\left[  -\frac{\hslash^{2}}{2}\frac{1}{\sqrt{-G}}\partial_{\Theta}\left(
\sqrt{-G}G^{\Theta\Pi}\partial_{\Pi}\right)  +U\left(  Q\right)  \right]
\Psi\left(  Q\right)  =0, \tag{3.1}%
\end{equation}
where $G_{\Theta\Pi}$ is a metric on the mini-superspace. In our case the
metric is given by (2.19) and $Q=\left(  a,r\right)  $. In the equation (3.1)
one has to take into account the factor ordering between the conjugate
variables. The kinetic terms of the Hamiltonian for $a$ and $r$ are obtained
from (2.17) and can be expressed in the equivalent forms on the classical
level:%
\[
\frac{P_{a}^{2}}{a}=a^{-\left(  j+k+1\right)  }P_{a}a^{j}P_{a}a^{k},
\]%
\[
\frac{P_{r}^{2}}{\Lambda T_{3}a^{3}H_{p}^{\left(  p-1\right)  /4}}=\frac
{1}{\Lambda T_{3}a^{3}}H_{p}^{-[\left(  p-1\right)  /4+u+w]}P_{r}H_{p}%
^{u}P_{r}H_{p}^{w},
\]
where $j,k,u,w$ are arbitrary numbers. On the quantum level they are
non-equivalent and depend on the commutation relations between the conjugate
variables. This ambiguity has to be respected in the quantum Hamiltonian. In
the canonical quantization the ordering ambiguity in the expression $f\left(
x\right)  P_{x}^{2}$ for the conjugated variables $\left(  x,P_{x}\right)  $
is resolved as follows:%
\begin{gather*}
f\left(  x\right)  P_{x}^{2}\rightarrow-\hslash^{2}\left(  1+A+B\right)
f\times\\
\left[  \frac{\partial^{2}}{\partial x^{2}}+\frac{A+2B}{1+A+B}\frac{f^{\prime
}}{f}\frac{\partial}{\partial x}+\frac{B}{1+A+B}\frac{f^{\prime\prime}}%
{f}\right]  ,
\end{gather*}
where $A,B$ are constants. Hence the differential part of the Hamiltonian
operator in our problem reads:%
\begin{gather*}
\frac{\hslash^{2}}{2m_{P}^{2}a}\left[  \frac{\partial^{2}}{\partial a^{2}%
}-\frac{\gamma}{a}\frac{\partial}{\partial a}+\frac{\delta}{a^{2}}\right]  +\\
-\frac{\hslash^{2}}{2\Lambda T_{3}a^{3}H_{p}^{\left(  p-1\right)  /4}}\left[
\frac{\partial^{2}}{\partial r^{2}}+\frac{\mu\left(  1-p\right)  }{4}%
\frac{H_{p}^{\prime}}{H_{p}}\frac{\partial}{\partial r}\right.  +\\
\left.  \frac{\nu\left(  p-1\right)  }{4H_{p}}\left(  \frac{p+3}{4}\frac
{H_{p}^{\prime}}{H_{p}}-H_{p}^{\prime\prime}\right)  \right]  ,
\end{gather*}
where $\gamma,\delta,\mu,\nu$ represent the factor ordering ambiguity and
prime denotes derivative with respect to $r.$ Thus the WD equation is:%
\begin{gather}
\left[  \frac{\partial^{2}}{\partial a^{2}}-\frac{\gamma}{a}\frac{\partial
}{\partial a}+\frac{\delta}{a^{2}}\right]  \Psi+\nonumber\\
-\frac{m_{P}^{2}}{\Lambda T_{3}a^{2}H_{p}^{\left(  p-1\right)  /4}}\left[
\frac{\partial^{2}}{\partial r^{2}}+\frac{\mu\left(  1-p\right)  }{4}%
\frac{H_{p}^{\prime}}{H_{p}}\frac{\partial}{\partial r}+\right. \nonumber\\
\left.  \frac{\nu\left(  p-1\right)  }{4H_{p}}\left(  \frac{p+3}{4}\frac
{H_{p}^{\prime}}{H_{p}}-H_{p}^{\prime\prime}\right)  \right]  \Psi+\nonumber\\
\frac{2m_{P}^{2}}{\hslash^{2}}U_{eff}\left(  a,r\right)  \Psi=0, \tag{3.2}%
\end{gather}
where an effective potential $U_{eff}\left(  a,r\right)  $ has the form:%
\begin{gather}
U_{eff}\left(  a,r\right)  =a^{4}\Lambda T_{3}\left(  H_{p}^{\left(
p-5\right)  /4}-2\Lambda\right)  -a^{2}\left(  m_{P}^{2}+\Lambda T_{3}\gamma
H_{p}^{\left(  p-5\right)  /4}\right)  +\nonumber\\
J^{2}a^{-2}H_{p}^{\left(  5-p\right)  /4}. \tag{3.2a}%
\end{gather}
At the moment we assume that $r$ is fixed and is considered as a parameter.
Thus the potential $U_{eff}$ and the wave function $\Psi$ have fixed values
for $r$ so the eq. (3.2) results in:%
\begin{equation}
\left[  \frac{d^{2}}{da^{2}}-\frac{\gamma}{a}\frac{d}{da}-\frac{\eta}{a^{2}%
}-\frac{2m_{P}^{4}k}{\hslash^{2}}a^{2}+\frac{2m_{P}^{2}\lambda}{3\hslash^{2}%
}a^{4}\right]  \Psi=0, \tag{3.3}%
\end{equation}
where $k$, $\lambda$ are given by (2.26-2.27) and $\eta=-C-\delta+\nu D$ with
$C$ and $D$ given by:%
\[
C=\frac{2m_{P}^{2}}{\hslash^{2}}J^{2}H_{p}^{\left(  5-p\right)  /4},
\]%
\[
D=\frac{m_{P}^{2}\left(  p-1\right)  }{4\Lambda T_{3}H_{p}^{\left(
p+7\right)  /4}}\left(  \left(  p+3\right)  H_{p}^{\prime}-4H_{p}H_{p}%
^{\prime\prime}\right)  .
\]
The last coefficient $D$ has the form:%
\begin{equation}
D\left(  r\right)  =-\frac{m_{P}^{2}\left(  p-1\right)  }{4\Lambda T_{3}%
H_{p}^{\left(  p+7\right)  /4}}g\left(  r\right)  \tag{3.4}%
\end{equation}
and:%
\begin{equation}
g\left(  r\right)  =\frac{4\left(  7-p\right)  \left(  8-p\right)  Ng_{s}%
}{r^{8-p}}[\frac{3+p}{4\left(  8-p\right)  }+\frac{1}{r}+\frac{Ng_{s}}%
{r^{8-p}}]. \tag{3.5}%
\end{equation}
The eq. (3.3) can be transformed to the following form:%
\begin{equation}
\frac{d^{2}F}{dz^{2}}+\frac{4s+1-\gamma}{2z}\frac{dF}{dz}+\left(
1-\frac{n^{2}}{z^{2}}+mz\right)  F=0, \tag{3.6}%
\end{equation}
where a variable $z$ is related to $a$ as follows:%
\[
z=-ia^{2}\frac{m_{P}^{2}}{\hslash}\sqrt{\frac{k}{2}},
\]
and the wave function $\Psi$ is expressed by a function $F$ in the form:%
\begin{equation}
\Psi\left(  z\right)  =z^{s}F\left(  z\right)  . \tag{3.7}%
\end{equation}
The coefficients $m$ and $n^{2}$ are given by:%

\[
m=-i\cdot\frac{\hslash\lambda}{3m_{P}^{4}k}\sqrt{\frac{2}{k}},\text{
\ \ }n^{2}=\frac{\eta-2s\left(  2s-\gamma-1\right)  }{4}.
\]
If we choose $s=\left(  1+\gamma\right)  /4$ in (3.6), then this equation
becomes:%
\begin{equation}
\frac{d^{2}F}{dz^{2}}+\frac{1}{z}\frac{dF}{dz}+\left(  1-\frac{n^{2}}{z^{2}%
}+mz\right)  F=0 \tag{3.8}%
\end{equation}
where:%
\begin{equation}
n^{2}=\eta/4+\left(  \gamma+1\right)  ^{2}/16 \tag{3.9}%
\end{equation}
This equation is similar to the Schr\"{o}dinger equation considered in the
Stark effect [10]. Because this equation has non analytical solutions the wave
function for the Stark effect is given only in a perturbation series in $m$.
First we consider the case when $m=0$ with a solution $F_{0}\left(  z\right)
$. The solution $F_{0}\left(  z\right)  $ corresponds to vanishing of the
cosmological constant $\lambda$ with the mirage perfect fluid on the
worldvolume. This fluid has the energy density $\rho=3J^{2}H_{p}^{\left(
5-p\right)  /4}/a^{6}$ and the state equation $w=+1$ (see eqs.(2.24) and
(2.28)). Under our assumption that $r$ is the parameter we are able to fix it
in such a way that $m$ is equal to zero. Thus one can relate $\Lambda$ to the
position $r_{0}$:%
\begin{equation}
2\Lambda=H_{p}^{\left(  p-5\right)  /4}\left(  r_{0}\right)  \tag{3.9}%
\end{equation}
and the parameter $k$ takes the form:%
\[
k=1+\frac{T_{3}\gamma}{2m_{P}^{2}}H_{p}^{\left(  p-5\right)  /2}\left(
r_{0}\right)  .
\]
Then for the fixed value of $r_{0}$ the equation (3.7) becomes the Bessel
equation:%
\begin{equation}
\frac{d^{2}F_{0}}{dz^{2}}+\frac{1}{z}\frac{dF_{0}}{dz}+\left(  1-\frac{n^{2}%
}{z^{2}}\right)  F_{0}=0. \tag{3.10}%
\end{equation}
The solution of (3.10) is given by:%
\[
F_{0}\left(  z\right)  =\widetilde{E}J_{n}\left(  z\right)  +\widetilde
{F}Y_{n}\left(  z\right)  ,
\]
where $J_{n}$ and $Y_{n}$ are the Bessel functions of the first kind of the
order $n$. Hence the wave function $\Psi$ obtained from (3.10) has the form:%
\begin{equation}
\Psi\left(  a;r_{0}\right)  =a^{\left(  1+\gamma\right)  /2}\left[
EI_{n}\left(  \frac{a^{2}}{l_{Pl}^{2}}\sqrt{\hslash^{2}k/2}\right)
+FK_{n}\left(  \frac{a^{2}}{l_{Pl}^{2}}\sqrt{\hslash^{2}k/2}\right)  \right]
, \tag{3.12}%
\end{equation}
where $E$ and $F$ are constants, $I_{n}\left(  z\right)  $ and $K_{n}\left(
z\right)  $ are the modified Bessel functions of the first kind and the second
kind. We introduced the Planck length $l_{Pl}=\hslash/m_{Pl}$. This wave
function must satisfy given boundary conditions. It has to be regular
everywhere, so using expansion of $I_{n}$ and $K_{n}$ near zero we obtain:%
\[
1+\gamma\geq\sqrt{4\eta+\left(  \gamma+1\right)  ^{2}}.
\]
It means that $4\eta\leq0,$ so in explicit form we get:%
\begin{equation}
-\frac{2m_{P}^{2}}{\hslash^{2}H_{p}^{\left(  p-5\right)  /4}\left(
r_{0}\right)  }J^{2}-\delta+\nu D\left(  r_{0}\right)  \leq0, \tag{3.13}%
\end{equation}
where $D\left(  r_{0}\right)  $ is given by the equation (3.4). From this
relation one can, in principle, determine the allowed positions of D3-brane in
the background for which the wave function of the D3-brane is given by (3.12).
For $z\rightarrow i\infty$ the asymptotics for $I_{n}$ and $K_{n}$ are
following:%
\[
I_{n}\left(  z\right)  \sim\frac{1}{\sqrt{2\pi z}}\exp\left(  z\right)  ,
\]%
\[
K_{n}\left(  z\right)  \sim\sqrt{\frac{\pi}{2z}}\exp\left(  -z\right)  .
\]
Thus the wave function (3.12) for big $a$ reads:%
\begin{gather}
\Psi\left(  a;r_{0}\right)  \simeq l_{Pl}\left(  \frac{2}{\hslash^{2}%
k}\right)  ^{1/4}a^{\left(  \gamma-1\right)  /2}\left[  e\exp\left(
+\frac{a^{2}}{l_{Pl}^{2}}\sqrt{\hslash^{2}k/2}\right)  +\right. \nonumber\\
\left.  f\exp\left(  -\frac{a^{2}}{l_{Pl}^{2}}\sqrt{\hslash^{2}k/2}\right)
\right]  . \tag{3.14}%
\end{gather}
The effects of the non-trivial D-brane background are included in $k$ and in
the order $n$ of the Bessel functions. The regularity of the wave function
near zero puts the constraint (3.13) on the positions of the D3-brane in the
background. The predictions following from this wave function are well-known
and broadly discussed with respect to the decay of a false vacuum [9].

The second exact solution of (3.6) is obtained for $s=\left(  \gamma-1\right)
/4$ and $n^{2}=0$. In this case (3.6) takes the form:%
\begin{equation}
\frac{d^{2}F}{dz^{2}}+\left(  1+mz\right)  F=0. \tag{3.15}%
\end{equation}
It is then transformed to the Airy equation:%
\begin{equation}
\frac{d^{2}\Phi}{dt^{2}}+\frac{1}{m^{2}}t\Phi=0, \tag{3.16}%
\end{equation}
where $t=1+mz$ and $F\left(  z\right)  =\Phi\left(  1+mz\right)  $. The
solutions $\Phi$ of the Airy equation depend on the sign of $m^{2}$. For
$m^{2}>0$ and for $m^{2}<0$ the solutions are given by:%
\begin{equation}
\Phi_{+}\left(  t\right)  =\frac{\sqrt{t}}{3}\left[  A^{\prime}J_{-1/3}\left(
\frac{2}{3|m|}t^{3/2}\right)  +B^{\prime}J_{1/3}\left(  \frac{2}{3|m|}%
t^{3/2}\right)  \right]  , \tag{3.17}%
\end{equation}%
\begin{equation}
\Phi_{-}\left(  t\right)  =\frac{\sqrt{t}}{3}\left[  AI_{-1/3}\left(  \frac
{2}{3|m|}t^{3/2}\right)  +BI_{1/3}\left(  \frac{2}{3|m|}t^{3/2}\right)
\right]  , \tag{3.18}%
\end{equation}
respectively. Since $m^{2}=-\left(  \frac{l_{Pl}^{4}\lambda}{6}\right)
^{2}\left(  \frac{2}{\hslash^{2}k}\right)  ^{3}<0$ the solution will be given
by the second relation. Finally the wave function $\Psi$ has the form:%
\begin{gather}
\Psi\left(  a;r_{0}\right)  =\frac{e^{i\pi s}}{3}\left(  \frac{1}{l_{Pl}^{2}%
}\sqrt{\frac{\hslash^{2}k}{2}}\right)  ^{s}a^{2s}\sqrt{1-\frac{\lambda
l_{Pl}^{4}}{3\hslash^{2}k}\frac{a^{2}}{l_{Pl}^{2}}}\times\nonumber\\
\times\left[  AI_{-1/3}\left(  \frac{4}{l_{Pl}^{4}\lambda}\left(
\frac{\hslash^{2}k}{2}\right)  ^{3/2}\left(  1-\frac{\lambda l_{Pl}^{4}%
}{3\hslash^{2}k}\frac{a^{2}}{l_{Pl}^{2}}\right)  ^{3/2}\right)  \right.
+\nonumber\\
\left.  BI_{1/3}\left(  \frac{4}{l_{Pl}^{4}\lambda}\left(  \frac{\hslash^{2}%
k}{2}\right)  ^{3/2}\left(  1-\frac{\lambda l_{Pl}^{4}}{3\hslash^{2}k}%
\frac{a^{2}}{l_{Pl}^{2}}\right)  ^{3/2}\right)  \right]  \tag{3.19}%
\end{gather}
and $s=\left(  \gamma-1\right)  /4$.\ This wave function is valid only for
$n^{2}=0$:%
\begin{equation}
\eta+\left(  \gamma-1\right)  \left(  \gamma+3\right)  /4=0. \tag{3.20}%
\end{equation}
The above condition fixed the position $r_{0}$ of D3-brane in the background
because $\eta$ is given by $H_{p}$ (see eqs. (3.4) and (3.5)). The wave
function (3.19) has the oscillating character for $3\hslash^{2}k/\left(
\lambda l_{Pl}^{2}\right)  <a^{2}$ and for all $a\geq0$ is regular. The (3.19)
has the form of the Hartle-Hawking wave function for empty universe with the
cosmological constant [11]. It is as should be since the condition (3.20)
implies that the worldvolume is empty with the cosmological constant $\lambda$.

Thus we have obtained two particular solutions of the eq. (3.6) which
correspond to the fixed positions of the D3-brane in the background. The first
solution (3.12) is obtained for vanishing of the cosmological constant
$\lambda$ on the worldvolume which is filled by the mirage fluid with the
energy density $3J^{2}H_{p}^{\left(  5-p\right)  /4}/a^{6}$. This solution is
valid only in the region where the condition (3.13) holds. The second solution
(3.19) is obtained for the position of D3-brane fixed by (3.20) and
corresponds to the empty worldvolume with the cosmological constant $\lambda$.

Next we consider the tunneling probability in the potential (3.2a) for a fixed
position $r_{0}$ of D3-brane with $m,n\neq0$. In order to do this we choose
$s=\left(  \gamma-1\right)  /4$\ in (3.6). Thus we get the one dimensional
equation:
\begin{equation}
\frac{d^{2}F}{dz^{2}}+\left(  1-\frac{n^{2}}{z^{2}}+mz\right)  F=0, \tag{3.21}%
\end{equation}
where:%
\begin{equation}
n^{2}=\eta/4+\left(  \gamma-1\right)  \left(  \gamma+3\right)  /16. \tag{3.22}%
\end{equation}
This equation can be considered as a Schr\"{o}dinger equation in a potential
$V\left(  z\right)  $ with a zero eigenvalue:%
\begin{equation}
\frac{d^{2}F}{dz^{2}}-V\left(  z\right)  F=0, \tag{3.23}%
\end{equation}
where $V\left(  z\right)  $ is given by:%
\begin{equation}
V\left(  z\right)  =-1+\frac{n^{2}}{z^{2}}-mz. \tag{3.24}%
\end{equation}
For the real variable $y=-\frac{a^{2}}{l_{Pl}^{2}}\sqrt{\hslash^{2}k/2}<0$ so
that $z=iy$ the potential $V$ takes the form
\begin{equation}
V\left(  y\right)  =-1-\frac{n^{2}}{y^{2}}+xn^{2}\left(  \frac{2}{\hslash
^{2}k}\right)  ^{3/2}y, \tag{3.25}%
\end{equation}
where $x=l_{Pl}^{4}\lambda/\left(  6n^{2}\right)  $ and $l_{Pl}$ is the Planck
length. This potential is sketched below:
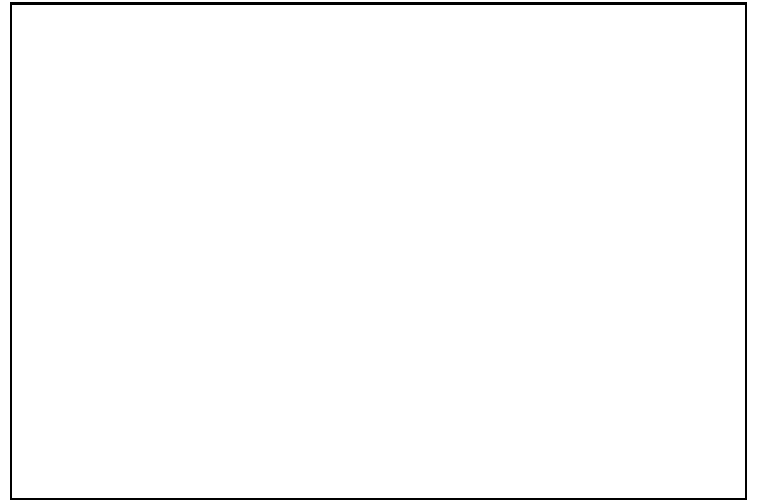
\begin{center}
\fbox{\includegraphics[
natheight=2.507100in,
natwidth=3.760200in,
height=2.5071in,
width=3.7602in
]%
{D:/SWP/graphics/PRdWdW__1.png}%
}
\end{center}
The variable $y$ has physical meaning only for $y<0$. The potential $V$ has a
maximum for
\[
y_{\max}=-\left(  \frac{2}{x}\right)  ^{1/3}\left(  \frac{\hslash^{2}k}%
{2}\right)  ^{1/2}%
\]
with the value:%
\begin{equation}
V_{\max}=-1-3n^{2}\left(  \frac{x}{2}\right)  ^{2/3}\frac{2}{\hslash^{2}k}.
\tag{3.26}%
\end{equation}
The scale factor $a_{0}^{2}$ which corresponds to $y_{\max}$ is:
\[
a_{0}^{2}=l_{Pl}^{2}\left(  \frac{2}{x}\right)  ^{1/3}%
\]
and is grater than $l_{Pl}$ for $x<2$.\ In the classical regime, if an energy
$E$ of the system is less then $V_{\max}$, we never pass from the region where
$a<a_{0}$ to the region where $a>a_{0}$. It means that the D3-brane collapse
to the zero size. But in the quantum regime exists non zero tunneling
probability to the region where $a>a_{0}$ which corresponds to expansion of
D3-brane worldvolume. This probability $\Gamma$ in the quasi-classical
approximation is given by:%
\begin{equation}
\Gamma\simeq\exp\left[  -\frac{2}{\hslash}\int_{y_{2}}^{y_{3}}\sqrt
{-E+V\left(  y\right)  }dy\right]  , \tag{3.27}%
\end{equation}
where the integration limits $y_{3}$, $y_{2}$ are given by the solution of the
equation:%
\[
E-V\left(  y\right)  =0.
\]
The roots $y_{3}$, $y_{2}$ are negative and ordered as follows: $y_{3}<$
$y_{2}<0$. Hence for $y\in\left(  y_{3},y_{2}\right)  $ and $E<V_{\max}<0$ we
obtain that:
\[
E-V\left(  y\right)  <0.
\]
Thus, an integral $I=\int_{y_{2}}^{y_{3}}\sqrt{-E+V\left(  y\right)  }dy$ is
real and has the form:%
\begin{equation}
I=n\sqrt{x}\left(  \frac{2}{\hslash^{2}k}\right)  ^{3/4}\int_{y_{2}}^{y_{3}%
}\frac{dy}{y}\sqrt{f\left(  y\right)  }, \tag{3.28}%
\end{equation}
a function $f$ is given by a cubic polynomial:%
\begin{equation}
f\left(  y\right)  =y^{3}+\sigma y^{2}-w, \tag{3.29}%
\end{equation}
where $\sigma=-w\left(  E+1\right)  /\left(  n^{2}\right)  $ and $w=\left(
\hslash^{2}k/2\right)  ^{3/2}/x$. Now we have to determine the integration
limits $y_{3}$and $y_{2}$ which depend on $E$. We make the reasonable
assumption (in the quasi-classical approximation) that the minimal value of
the scale factor $a$ on the worldvolume is given by the Planck length
$l_{Pl}=\hslash/m_{Pl}$. This minimal value corresponds to $y_{Pl}%
=-\hslash\sqrt{k/2}$. At this point the potential $V$ has the value:%
\begin{equation}
V\left(  y_{Pl}\right)  =-1-2n^{2}\frac{1+x}{\hslash^{2}k}, \tag{3.30}%
\end{equation}
and $V_{\max}>V\left(  y_{Pl}\right)  $. The coefficient $\sigma$ for
$E=V\left(  y_{Pl}\right)  $ takes the form:%
\begin{equation}
\sigma=\sqrt{\frac{\hslash^{2}k}{2}}\left(  1+\frac{1}{x}\right)  . \tag{3.31}%
\end{equation}
We evaluate $\Gamma$ near the energy $E=V\left(  y_{Pl}\right)  $. Thus:%
\begin{gather}
\sqrt{-E+V\left(  y\right)  }=\sqrt{-V\left(  y_{Pl}\right)  +V\left(
y\right)  }=\nonumber\\
=n\sqrt{x}\left(  \frac{2}{\hslash^{2}k}\right)  ^{3/4}\frac{1}{y}%
\sqrt{f\left(  y\right)  }. \tag{3.32}%
\end{gather}
It means that the cubic polynomial $f\left(  y\right)  $ has a root for
$y=y_{Pl}<0$ and the bottom integration limit is $y_{2}=y_{Pl}$. The other
roots are given by:%
\begin{equation}
y_{1}=\frac{1}{2x}\sqrt{\frac{\hslash^{2}k}{2}}\left(  -1+\sqrt{1+4x}\right)
, \tag{3.33}%
\end{equation}%
\begin{equation}
y_{3}=-\frac{1}{2x}\sqrt{\frac{\hslash^{2}k}{2}}\left(  1+\sqrt{1+4x}\right)
\tag{3.34}%
\end{equation}
and they are ordered in the following way $y_{3}<y_{Pl}<0$ and $y_{1}>0$. The
upper integration limit is $y_{3}$ and the polynomial $f$ $=(y-y_{3})\left(
y-y_{Pl}\right)  \left(  y-y_{1}\right)  $ is positive for $y\in\left(
y_{3},y_{Pl}\right)  $. The barrier width $b$ which is equal to $y_{Pl}-y_{3}%
$:%
\begin{equation}
b=\sqrt{\frac{\hslash^{2}\kappa}{2}}\left(  -1+\frac{1+\sqrt{1+4x}}%
{2x}\right)  \tag{3.35}%
\end{equation}
is greater then zero for $x\in\left(  0,2\right)  $ and vanishes for $x=2$.

The integral (3.28) is the elliptic integral and is expressed as follows:%
\begin{equation}
I=n\sqrt{x}\left(  \frac{2}{\hslash^{2}k}\right)  ^{3/4}\left(  \frac{\sigma
}{3}I_{1}-wI_{-1}\right)  , \tag{3.36}%
\end{equation}
where:%
\begin{equation}
I_{s}=\int_{y_{Pl}}^{y_{3}}\frac{y^{s}}{\sqrt{f\left(  y\right)  }}dy
\tag{3.37}%
\end{equation}
and $s=-1,1$. Making the standard substitution $y=y_{3}+\left(  y_{Pl}%
-y_{3}\right)  \sin^{2}\phi$ in the integrals $I_{s}$ we obtain:%
\begin{equation}
I_{1}=-\frac{2y_{1}}{\sqrt{y_{1}-y_{3}}}K\left(  \kappa\right)  +2\sqrt
{y_{1}-y_{3}}E\left(  \kappa\right)  , \tag{3.38}%
\end{equation}%
\begin{equation}
I_{-1}=-\frac{2}{y_{3}\sqrt{y_{1}-y_{3}}}\Pi\left(  c,\kappa\right)  ,
\tag{3.39}%
\end{equation}
where $K\left(  \kappa\right)  ,$ $E\left(  \kappa\right)  $ and $\Pi\left(
c,\kappa\right)  $ are the first kind, the second kind and the third kind
complete elliptic integrals, respectively (see Appendix). The modulus $\kappa$
and $c$ are given by:%
\begin{equation}
\kappa=\frac{y_{Pl}-y_{3}}{y_{1}-y_{3}}=\frac{1}{2}\left(  1+\frac{1-2x}%
{\sqrt{1+4x}}\right)  , \tag{3.40}%
\end{equation}%
\begin{equation}
c=-\frac{y_{Pl}-y_{3}}{y_{3}}=\frac{1}{2}\left(  3-\sqrt{1+4x}\right)  >0.
\tag{3.41}%
\end{equation}
They have the range: $0<\kappa<1$ and $0<c<1$ for $x\in\left(  0,2\right)  $.
Thus the integral $I$ takes the form:%
\begin{gather}
I=n\sqrt{x}\left(  \frac{2}{\hslash^{2}k}\right)  ^{3/4}\left(  -\frac{2\sigma
y_{1}}{3\sqrt{y_{31}}}K\left(  \kappa\right)  +\frac{2\sigma\sqrt{y_{31}}}%
{3}E\left(  \kappa\right)  +\right. \nonumber\\
\left.  \frac{2w}{y_{3}\sqrt{y_{1}-y_{3}}}\Pi\left(  c,\kappa\right)  \right)
, \tag{3.42}%
\end{gather}
where $y_{31}=y_{1}-y_{3}>0$ and $y_{32}=y_{2}-y_{3}>0$. Collecting the above
formulas we get the function $I$ in the variables $x$, and $n$:%
\begin{gather}
I\left(  x,n\right)  =n\frac{1-\sqrt{1+4x}}{\left(  1+4x\right)  ^{1/4}}%
\Pi\left(  c,\kappa\right)  +\nonumber\\
n\frac{1+x}{6x\left(  1+4x\right)  ^{1/4}}\left[  4\sqrt{1+4x}E\left(
\kappa\right)  +\left(  1-\sqrt{1+4x}\right)  K\left(  \kappa\right)  \right]
. \tag{3.43}%
\end{gather}
This function tends to $\infty$ if $x\rightarrow0$ and is decreasing function
of $x$ with the minimum for $x=2$
\begin{equation}
I\left(  2,n\right)  =\frac{\pi n}{4\sqrt{2}}. \tag{3.44}%
\end{equation}
For the value of $x=2$ the barrier width $b$ vanishes and $n^{2}=l_{Pl}%
^{4}\lambda/12$. The maximum of the tunneling probability $\Gamma$ is given
by:%
\begin{equation}
\Gamma_{\max}=\exp(-\frac{2}{\hslash}I\left(  2,n\right)  )=\exp\left(
-\frac{\pi n}{2\sqrt{2}\hslash}\right)  . \tag{3.45}%
\end{equation}
Our result agrees with [12] in the sense that maximum of the probability is
achieved in the point where the barrier potential vanishes. Moreover $n$ is
the function of $r$ and parameters related to the ordering (see eqs. (3.22),
(3.4) and (3.5)). Thus $\Gamma_{\max}$ takes the maximum for $n=0$ with the
cosmological constant $\lambda=0$ on the world-volume. This case corresponds
to the position $r_{\max}$ for which:%
\[
n\left(  r_{\max}\right)  =0.
\]
Thus $r_{\max}$ gives the boundary value of the condition (3.13). In this
position the D3-brane has the wave function given by (3.12).

\section{Conclusions}

We have considered the gravity on the worldvolume of D3-brane embedded in the
flat background produced by $N$ $p$-branes. This system is described by the
sum of the Hilbert-Einstein and DBI actions. Although the DBI action is
non-linear we have used equivalent form for it with an auxiliary metric. This
metric has been promoted to the dynamic field on the worldvolume. Thus the
worldvolume gravity is described by this metric. In the classical regime we
have obtained potential (2.18). Comparison of the classical equations of
motion for this potential (for fixed $r$) to the Friedmann equations results
that (2.20) and (2.21) can be interpreted as the energy density and the
pressure of the mirage perfect fluid. For the big value of the scale factor
$a$ we get the state equation of this perfect fluid in the form (2.27). Thus
one can conclude that the expansion of the worldvolume is accelerated.

In the next part of this paper we have derived the Wheeler-de Witt equation
for this system and solved it for the special cases corresponding to the fixed
positions of $D3$-brane in the background. In the first case the wave function
has the form of the instanton tunneling and in the second we get the
Hartle-Hawking function. The obtained wave functions depend on the ordering of
the conjugated variables. In general, the wave function depends on $a$ and $r$
but we can not find it in the explicit form. Even for fixed $r $ we get an
equation which has not explicit solutions and is similar to the equation
considered in the Stark effect. The perturbation method used in the Stark
effect is inappropriate because we do not know which parameter is small in eq.
(3.8). We made the assumption that the minimal value of the scale factor $a$
of the worldvolume is given by the Planck length and for this value we
evaluated tunneling probability in the potential $V$. This probability is
finite and depends on $\lambda$ and $r$. The maximum of the probability is for
$x=2$ and the cosmological constant is given by $\lambda=12n^{2}l_{Pl}^{-4}$.

The considered model has the common features with a hydrogen atom in an
external electric field. In our model the external field is represented by the
background produced by Dp-branes and the atom is represented by D3-brane. The
tunneling from Planckian size of D3-brane to classical regime corresponds to
the ionization of the hydrogen atom.

The above model is very simple and not quite realistic for several reasons.
The first one is that we used FRW ansatz for the metric. Validity of this
metric was extended to the pre-inflation epoch, when quantum gravity has
dominating effects. In this regime the universe was not necessary isotropic
and homogenous. The better approximation for the metric is IX Bianchi type.
This case were considered in [13] for the empty universe with a cosmological
constant. The second reason is\ that the stringy effects enter by the DBI
action. This action is valid only on the tree level of the string theory and
does not take into account backreactions. The other reasons are related to the
special backgrounds which are classical solutions of the supergravity
equations. The more general approach is if one consider\ the Wheeler-De Witt
equation for the low-energetic approximation of string theory by supergravity
and next compactificate this equation to 4-dimensions taking into account
stringy corrections.

In the classical regime compactification to 4-dimensions for the Bianchi
type-I cosmology in the presence the gravity, the dilaton, and the
antisymmetric tensor field of the second rank, coupling to the gauge field
strength living on the D3-brane was considered in [14]. It would be
interesting to consider the quantum regime of this system, described \ by\ the
Wheeler-DeWitt equation, and compare the results with the results of the model
considered in this paper. We shall investigate this problem in the future.

\section{Appendix}

The complete elliptic integrals first kind $K$, second kind $E$ and third kind
$\Pi$ are defined as follows [15]:%
\[
K\left(  \kappa\right)  =\int_{0}^{\pi/2}\frac{d\phi}{\sqrt{1-\kappa\sin
^{2}\phi}},
\]%
\[
E\left(  \kappa\right)  =\int_{0}^{\pi/2}\sqrt{1-\kappa\sin^{2}\phi}d\phi,
\]%
\[
\Pi\left(  c,\kappa\right)  =\int_{0}^{\pi/2}\frac{d\phi}{\left(  1-c\sin
^{2}\phi\right)  \sqrt{1-\kappa\sin^{2}\phi}}.
\]
For $c=\kappa$ the elliptic integral $\Pi$ is expressed by $E$:%
\[
\Pi\left(  \kappa,\kappa\right)  =\frac{E\left(  \kappa\right)  }{1-\kappa}.
\]

\section{References}

[1] M. Alishahiha, E. Silverstein, D. Tong, \textit{DBI in the sky}, Phys.
Rev. D 70, 123505 (2004) [arXiv:hep-th/0404084]; E. Silverstein, D. Tong,
\textit{Scalar speed limits and cosmology: Acceleration from D-cceleration},
Phys. Rev. D 70, 103505 (2004) [arXiv:hep-th/0310221].

[2] M.Abou-Zeid, C. M. Hull, \textit{Geometric Actions for D-Branes and
M-Branes}, Phys. Lett. B428 (1998) 277, [hep-th/9802179] ; M. Abou Zeid,
\textit{Actions for Curved Branes}, hep-th/0001127.

[3] U. Lindstr\"{o}m, R. von Unge, \textit{A Picture of D-branes at Strong
Coupling}, Phys.Lett. B403 (1997) 233-238, hep-th/9704051

[4] R. Kallosh, \textit{Covariant quantization of D-branes}, Phys.Rev. D56
(1997) 3515-3522, hep-th/9705056

[5] K. Kamimura and M. Hatsuda, \textit{Canonical formulation of IIB
D-branes}, Nucl. Phys. B527 (1998) 381, hep-th/9712068

[6] S. Kachru, R. Kallosh, A. Linde, S. P. Trivedi, \textit{de Sitter Vacua in
String Theory}, [arXiv:hep-th/0301240].

[7] G. T. Horowitz, A. Strominger, \textit{Black strings and P-branes}, Nucl.
Phys. \textbf{B360} (1991) 197

[8] T. Tanaka, M. Sasaki, K. Yamamoto, \textit{Field-theoretical description
of quantum fluctuations in the multidimensional tunneling approach}, Phys.
Rev. D49, (1994) 1039

[9] U. Gen, M. Sasaki, \textit{False Vacuum Decay with Gravity in
Non-Thin-Wall Limit}, Phys.Rev. D61 (2000) 103508, gr-qc/9912096

[10] E. M. Lifshitz, L. D. Landau, \textit{Quantum Mechanics}, Pergamon Press,
Oxford (1965)

[11] J. B. Hartle, S. W. Hawking, \textit{Wave function of the universe,}
Phys. Rev.D28 (1983) 2960

[12] R. Brunstein, S. P. de Alvis, \textit{The Landscape of String Theory and
The Wave Function of the Universe,} Phys.Rev. D73 (2006) 046009 , hep-th/0511093

[13] R. Graham, P. Szepfalusy, \textit{Quantum creation of a generic
universe}, Phys. Rev. D42 (1990) 2483

[14] I. Cho, E. J. Chun, H. B. Kim, Y. Kim, \textit{Cosmological Aspects of
the D-brane World,} hep-th/0607040

[15] M. Abramowitz, I. A. Stegun, \textit{Handbook of Mathematical Functions},
(1964) Dover Publications, New York.
\begin{verbatim}

\end{verbatim}

\end{document}